\documentstyle[hep93]{article}
\input epsf
\newcommand{\CP}{\mbox{\it CP\,}}
\newcommand{\CPT}{\mbox{\it CPT\,}}
\newcommand{\eg}{{\em e.g.\ }}
\newcommand{\ie}{{\em i.e.\ }}
\newcommand{\vs}{{\em vs.\ }}
\newcommand{\cf}{{\em cf.\ }}

\begin{document}
\title {
\begin{flushright}
{\normalsize IIT-HEP-96/2\\
\vspace{ -.1in}
hep-ex/9509009
}
\end{flushright}
\vspace{0.1in}
{\bf Fixed-Target {\em CP}-Violation Experiments at Fermilab}\thanks{Invited 
talk presented at
the 5th Hellenic School and Workshops
on Elementary Particle Physics, Corfu, Greece, 3--24 Sept.\ 
1995.}\thanks{This is a
somewhat revised and expanded version of a review to appear in {\sl Proc.\ Four
Seas Conference}, Trieste, Italy, 25 June -- 1 July 1995.}
}
\author{Daniel M. Kaplan\thanks{E-mail: kaplan@fnal.gov}\\
{\sl Illinois Institute of Technology, Chicago IL 60616, USA} \\
{\rm (representing the HyperCP and Charm2000 collaborations)}
}
\abstract{
Studies of {\CP} violation, for 30 years focused primarily on the neutral $K$ 
meson, are on the threshold of a new era as experiments approach 
Standard-Model sensitivities in decays of beauty, charm, and hyperons. The 
array of heavy-quark experiments approved and planned at Fermilab may lead to 
a significant breakthrough in the next five to ten years.
}

\date{}

\maketitle

\section{Introduction}

The asymmetry of certain weak decays with respect to the simultaneous 
interchange of particles with antiparticles ($C\,$) and reflection of 
spatial coordinates ($P\,$)~\cite{Cronin-Fitch} raises 
fundamental questions about space, time, and the early history of the 
Universe~\cite{Sakharov}. Despite thirty years of impressive experimental 
effort, we still have little insight into the origin of this phenomenon. New 
experimental approaches now being attempted may lead to substantially improved 
understanding in the next five to ten years.

{\CP} violation can most simply be thought of as a difference in decay
properties between particles and antiparticles. For such a difference to arise,
there must be competing decay amplitudes which interfere, leading to a phase
difference whose magnitude changes under {\em CP} transformation. Since there
is no evidence for {\CP} asymmetry in strong-interaction or electromagnetic 
processes, it is generally assumed that this interference arises in the weak
sector. 

\begin{figure}[htb]
\vspace{0.in}
\centerline{\hspace{-0.625in}\epsfxsize = 4 in \epsffile {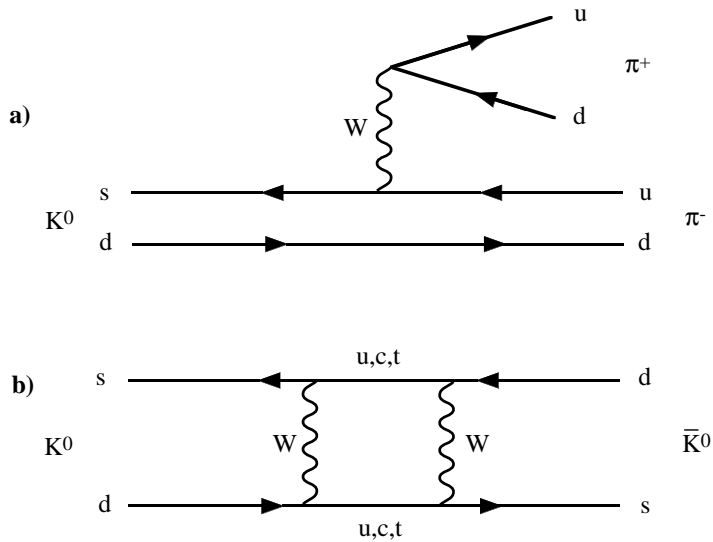}}
%\vspace{-0.5in}
\caption {Examples of a) direct $K^0$ decay and b) $K^0$ mixing via a ``box"
diagram.
\label{K0decay}}
\end{figure}

The prototypical example of {\em CP} violation is that arising from
particle-antiparticle mixing in the neutral-kaon system. The two processes
that interfere in this case are the direct decay of the $K^0$
(Fig.~\ref{K0decay}a) and decay
occurring after conversion (through mixing) into $\overline {K^0}$
(Fig.~\ref{K0decay}b). 
As a result, the physical $K_S$ and $K_L$ states are not {\em CP} eigenstates 
(discussed in more detail in Section~\ref{Direct} below), thus $K_L$ (and 
presumably also $K_S$) can decay
into both {\em CP}-odd and {\em CP}-even
final states~\cite{Cronin-Fitch}.
As first
pointed out by Kobayashi and Maskawa~\cite{KM}, 
in a six-quark model the participation of all three
quark generations in the mixing process introduces a nontrivial weak phase in
the mixing amplitude, which changes sign under {\em CP}. Since the
amplitudes for the direct and mixed decays 
can also possess a strong phase difference which is {\em CP}-invariant, the
combined phase difference can change in magnitude under {\em CP}.

In the Standard Model (SM), {\CP} violation at possibly-observable levels 
can occur in the decays of neutral~\cite{Heinrich,K0CP} and charged~\cite{KCP} 
kaons, hyperons~\cite{HyperCP}, and charm~\cite{CharmCP} and 
beauty~\cite{Rosner} mesons. To date it has been observed in only the first of 
these cases. The pattern of occurrence in all of them could reveal whether 
{\CP} violation originates (as in the SM) solely from the one irreducible phase
of the Cabibbo-Kobayashi-Maskawa (CKM) quark-mixing
matrix~\cite{Cabibbo,KM}, whether in
addition there are contributions from new physics outside the CKM framework, or
whether the phenomenon arises entirely from new physics. For new-physics
contributions, {\CP} violation probes multi-TeV mass scales which cannot be
studied directly even at the LHC. 

Given the sizes of {\em CP} asymmetries observed in $K^0$ decays and the values 
of the CKM matrix elements, the SM predicts a distinct hierarchy of
{\CP}-violating effects:  {\em CP} asymmetries should be largest
($_\sim$\llap{$^>$}10$^{-1}$) in certain relatively
rare beauty decays, smaller in decays of $K^0$ (few$\,\times10^{-3}$)
and Cabibbo-suppressed charm decays ($\sim10^{-3}$), and smaller 
still ($\sim10^{-5}-10^{-4}$) in decays of hyperons.
Nevertheless, the sizes of production cross sections and branching ratios make 
detection of these effects 
hardest in principle in the beauty sector and easiest in the kaon 
sector, with hyperons and charm lying in between.
While until recently, SM charm and hyperon {\CP} asymmetries appeared beyond 
reach, advances in data-acquisition technology have now made their observation 
feasible, and experiments are now being mounted or proposed to 
search for them. 
Observation of {\em CP} asymmetries in beauty decay requires construction of new
accelerators~\cite{Nozaki,PEP-II}, ambitious new experiments~\cite{HERA-B}, or
substantial upgrades to existing
accelerators or experiments~\cite{DeJongh,Berkelman}; 
all of these efforts are also in progress.

Charm studies can play a special role because
the top-quark loops which in the SM dominate {\CP} violation in the
strange and beauty sectors are absent, 
creating a low-background window for new physics,
and because new physics may couple
differently to up- and down-type quarks or couple to quark mass. 
If kaon and beauty experiments confirm the CKM model, we will be 
hardly any closer to an ultimate theory of {\em CP} violation, since the 
question why the CKM phase has the value it does will remain open.
On the other hand, by pursuing this physics in all available quark sectors,
we may find deviations from CKM predictions which could point the way to
a deeper understanding.
Many of these
issues are treated in more detail in the excellent recent reviews of Winstein
and Wolfenstein~\cite{Winstein-Wolfenstein} and Rosner~\cite{Rosner}; a more
detailed discussion of hyperon {\CP} violation can be found in the Fermilab
Experiment 871 Proposal~\cite{E871}.

As suggested by Table~\ref{tab:FNAL}, Fermilab fixed-target experiments have
made substantial contributions to this subject in recent years and will
continue to do so in the years ahead. At Fermilab the search for {\CP} 
violation in beauty decay is part of the Tevatron Collider  program and will 
not be pursued in fixed target. The remainder of this article therefore 
reviews the kaon, hyperon, and charm programs at Fermilab.

\begin{table}[htb]
\centering
\caption
{Recent and future Fermilab fixed-target {\CP}-violation experiments (question 
marks designate experiments not yet approved).}
\label{tab:FNAL}
\vspace{5mm}
%\footnotesize
%
\begin{tabular}{|lcccc|}
\hline
Run:  & 1987/8 &1990/1 & 1996/7 & $_\sim$\llap{$^>$}2000 \\
\hline
\hline
\multicolumn {5}{|l|}{Kaon experiments:}\\
\hline
& E731 & E773/E799-I & KTeV & KTeV/KAMI? \\
\hline
\multicolumn {5}{|l|}{Hyperon experiments:}\\
\hline
& E756 & & HyperCP & \\
\hline
\multicolumn {5}{|l|}{Charm experiments:}\\
\hline
& & E687 & FOCUS & Charm2000? \\
& & E791 & & \\
\hline
\end{tabular}
\end{table}

\section{The Search for Direct {\CP} Violation in $K^0$ Decay}
\label{Direct}
\begin{figure}[htb]
\vspace{0.in}
\centerline{\hspace{-0.5in}\epsfxsize = 4 in \epsffile {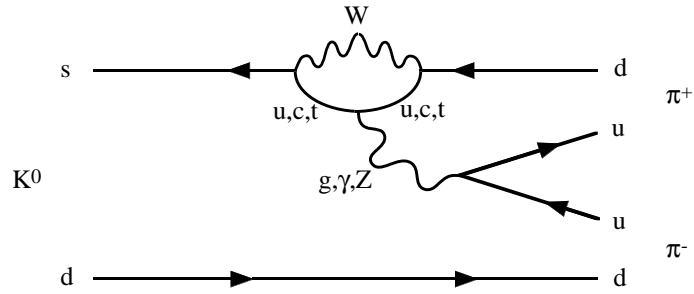}}
%\vspace{-0.5in}
\caption {Example of $K^0$ decay via a ``penguin" diagram.
\label{penguin}}
\end{figure}

A question that has received much attention is whether all {\CP} violation 
arises indirectly (as predicted in the ``superweak" theory of
Wolfenstein~\cite{superweak}), \ie through the mixing of neutral mesons
with their 
antiparticles, or whether there is in addition direct {\CP} violation,
arising in the decay process itself. 
While only indirect {\CP} violation has so far been observed, the Standard 
Model also predicts observable levels of direct {\CP} violation, arising 
from the interference of ``penguin" diagrams~\cite{Ellis} 
(containing $W$ loops, see Fig.~\ref{penguin})
with tree-level diagrams. To date 
the search for direct {\CP} violation has mainly concentrated on the 
measurement of the ratio $\epsilon^\prime/\epsilon$, where $\epsilon$ 
parametrizes the degree to which $K_S$ and $K_L$ are not {\CP} eigenstates,
\begin{eqnarray}
|K_S\rangle &=& [(1+\epsilon)|K^0\rangle + (1-\epsilon)|\overline {
K^0}\rangle]/\sqrt{2(1+|\epsilon|^2)} \\ 
|K_L\rangle &=& [(1+\epsilon)|K^0\rangle - (1-\epsilon)|\overline {
K^0}\rangle]/\sqrt{2(1+|\epsilon|^2)}\,, 
\end{eqnarray}
and $\epsilon^\prime$ measures the difference in {\CP}-violating decay rates 
of $K_L$ to $\pi^+\pi^-$ and $\pi^0\pi^0$:
\begin{eqnarray}
\frac{\epsilon^\prime}{\epsilon}&=
&\frac{1}{6}\bigg(1-\bigg|\frac{\eta_{00}}
{\eta_{+-}}\bigg|^2\bigg)\\ 
&\equiv&\frac{1}{6}\bigg[1-\frac{\Gamma(K_L\to\pi^0\pi^0)/
\Gamma(K_S\to\pi^0\pi^0)}{
\Gamma(K_L \to\pi^+\pi^-)/\Gamma(K_S\to\pi^+\pi^-)}\bigg]\,.
\end{eqnarray}
A nonzero value of $\epsilon^\prime/\epsilon$ indicates {\CP} 
violation in $\Delta S=1$ $K^0$ decays, and not solely through $\Delta S= 2$ 
mixing. The measurement of $\epsilon^\prime/\epsilon$ entails 
determination of four decay rates, which can be carried out such that 
systematic 
uncertainties cancel in the double ratio 
$\frac{\Gamma(K_L\to\pi^0\pi^0)/\Gamma(K_S\to\pi^0\pi^0)}{\Gamma(K_L
\to\pi^+\pi^-)/\Gamma(K_S\to\pi^+\pi^-)}$. In this way sensitivity at the 
$_\sim$\llap{$^<$}\,$10^{-4}$ level can be achieved~\cite{Winstein-Wolfenstein}.

The SM expectation for $\epsilon^\prime/\epsilon$ is sensitive to the 
value of the top-quark mass, because of competition between the strong and 
electroweak penguin diagrams, which contribute with opposite 
signs~\cite{Flynn_Buchalla}. The degree 
of this sensitivity is unsettled in the literature. Some 
authors find complete cancellation, $\epsilon^\prime/\epsilon$ 
becoming zero at $m_t \approx 220\,$GeV and negative  for larger top 
mass~\cite{K0CP,Paschos}. But 
Heinrich {\it et al.}~\cite{Heinrich},
 using chiral perturbation theory, find the 
cancellation to be only partial, $\epsilon^\prime/\epsilon$ remaining 
positive for all $m_t$ values. Thus at the present state of understanding, and 
given $m_t= 180 \pm 12\,$GeV~\cite{mtop}, it appears that the SM predicts 
$\epsilon^\prime/\epsilon$ in the range (0 to 
3)$\times10^{-3}$~\cite{Heinrich,Paschos2};
some maintain that it 
cannot exceed $1\times10^{-3}$~\cite{K0CP,Rosner}.
This range will presumably decrease 
with further  theoretical effort and improvement in the determination of $m_t$.

$\epsilon^\prime/\epsilon$ has also been estimated in a variety of 
extensions of the SM~\cite{Donoghue,Winstein-Wolfenstein,Rosner}. 
It is an old idea that {\CP} violation may originate through spontaneous 
symmetry breaking in the Higgs sector~\cite{Lee}. In Weinberg's 
multiple-Higgs-doublet model~\cite{Weinberg}, assuming that Higgs exchange is 
a major contributor to $\epsilon$, $\epsilon^\prime/\epsilon$ can be as large 
as $\cal{O}$$(10^{-2})$~\cite{Donoghue,Winstein-Wolfenstein, Bigi-Sanda}, and 
an electric dipole moment for the neutron $d_n\, 
{}_\sim$\llap{$^>$}\,10$^{-25}\,e$\,cm is also predicted~\cite{Bigi-Sanda}. 
Given the current experimental limit $d_n 
<1.1\times10^{-25}\,e$\,cm~\cite{PDG}, the Weinberg model may still be viable, 
however a substantial lowering of $d_n$, or establishment of a sufficiently 
small value for $\epsilon^\prime/\epsilon$, could rule out this model as a 
significant source of $\epsilon$~\cite{Bigi-Sanda,Winstein-Wolfenstein}. 
Alternative multi-Higgs models have also been formulated~\cite{tree}, in which 
the ``natural flavor conservation"~\cite{Weinberg-Glashow} of the Weinberg 
model is abandoned in favor of an approximate family 
symmetry~\cite{Bigi-Sanda, Hall-Weinberg,Wu}. In these models, if {\CP} 
violation is attributed to flavor-changing neutral-Higgs exchange (FCNE), all 
direct {\CP} violation (and all {\CP}-violating effects in beauty) can be 
unobservably small, but there are other observable manifestations, such as 
large mixing in charm~\cite{Hall-Weinberg}
(see Section~\ref{sec:mixing} below). The general analysis of Wu and 
Wolfenstein~\cite{Wu} includes {\CP}-violating charged-Higgs exchange, leading 
to a richer variety of possibilities; for example, $\epsilon^\prime/\epsilon$ 
can then be as large as in the CKM model.
Minimal supersymmetry (despite having an extra Higgs doublet)
predicts zero for $\epsilon^\prime/\epsilon$
due to the relatively real vacuum expectation values of the two
doublets~\cite{Rosner}. Left-right-symmetric models, featuring extra 
right-handed gauge bosons with masses well above those of the left-handed 
ones, seek to provide a unified explanation of $P\,$ and {\CP} violation in 
which both symmetries are conserved at sufficiently high energy but 
spontaneously broken at low energy~\cite{Mohapatra}.
``Isoconjugate" left-right models~\cite{Mohapatra-Pati} predict zero for 
$\epsilon^\prime/\epsilon$~\cite{Donoghue, Mohapatra}, but other versions can 
accommodate values as large as
$5\times10^{-3}$~\cite{Winstein-Wolfenstein,Mohapatra}. In models with 
appreciable left-right mixing, $\epsilon^\prime$ and $d_n$ become 
related~\cite{Mohapatra}: $\epsilon^\prime/d_n\simeq10^{21}\,(e\,{\rm 
cm})^{-1}$.

The experimental situation is as follows.  Two experiments, one (E731)
at Fermilab  and one 
(NA31) at CERN, have published results with comparable sensitivity which are 
1.8$\sigma$ 
apart. E731 obtains Re$(\epsilon^\prime/\epsilon) = (7.4\pm5.2\pm2.9)
\times10^{-4}$~\cite{Gibbons}, where the first error is statistical and the 
second systematic, while the NA31 result is
Re$(\epsilon^\prime/\epsilon)=(23\pm3.6\pm5.4)\times10^{-4}$~\cite{Barr}. 
Averaging these with previous results
from the Fermilab collaboration~\cite{Woods}, the Particle Data Group 
finds Re$(\epsilon^\prime/\epsilon)=(1.5\pm0.8)\times10^{-3}$~\cite{PDG}, 
employing their standard procedure for increasing the uncertainty to take 
account of the NA31--E731 disagreement. While the NA31 
result is $3\sigma$ from zero, the world average is less than $2\sigma$ from 
zero, 
thus we cannot conclude that direct {\CP} violation has been observed. 

The techniques employed by the two groups differ in important ways. For 
example, in E731 two parallel $K_L$ beams were incident, and a regenerator 
placed in one created a $K_S$ beam at the upstream end of the decay region.
In NA31 a $K_S$ production target was moved throughout the decay region to 
minimize acceptance differences for $K_L$ and $K_S$ decays. E731 used 
magnetic spectrometry for the final-state charged pions and lead-glass 
calorimetry for neutrals, while NA31 relied on liquid-argon 
calorimetry for energy measurement in all modes.
While in E731 both $K_L$ and $K_S$ decays were acquired simultaneously, in NA31 
$\pi^+\pi^-$ and $\pi^0\pi^0$ final states were acquired simultaneously, thus 
temporal variations in operating conditions had differing effects in the two 
experiments.
Both 
groups are preparing improved experiments, designated KTeV (Fermilab) and NA48 
(CERN). Since the E731 uncertainty is dominated by statistical error, the 
Fermilab collaboration has elected to retain the E731 approach with an upgraded 
apparatus~\cite{KTeV}. NA48, however, represents a substantial departure from 
NA31, for 
example adopting the technique of magnetic momentum 
analysis for the charged-pion final state~\cite{NA48}. 
In the new experiments, 
both groups intend to take all four modes simultaneously.
The goal for each effort is 
sensitivity of $(1-2)\times10^{-4}$~\cite{Winstein-Wolfenstein}.

\section{Other $K^0$ Studies}

The E731 collaboration has also performed a sensitive test of {\CPT} 
symmetry in $K^0$ decay. In E773 they modified their regenerator arrangement so
as to make a precise measurement of the phases of $\eta_{+-}$ and $\eta_{00}$.
{\CPT} symmetry predicts these phases to be equal and also relates their
size to $\Delta m_K$ and $\Delta\Gamma_K$~\cite{Barmin}. 
The E773 results, 
$\phi_{00}-\phi_{+-} = 0.62^\circ \pm 1.03^\circ$ and $\phi_{+-}=43.53^\circ
\pm 0.97^\circ$~\cite{Schwingenheuer}, confirm the predictions and
are the most precise {\CPT} tests to date, improving on 
previous results from E731~\cite{GibbonsCPT}.

Direct {\CP} violation can also be sought in rare decays of $K^0$.  
The decay rates for $K_L\to\pi^0e^+e^-$, $\pi^0\mu^+\mu^-$, and
$\pi^0\nu\overline{\nu}$ are expected to be dominated by direct {\CP}-violating
processes~\cite{Donoghue,Winstein-Wolfenstein,Rosner}. (In the first two cases 
there are also {\CP}-conserving contributions occurring via virtual-photon 
loops, which are monitored by $K_L\to\pi^0\gamma\gamma$~\cite{Donoghue}.) 
In E799-I, 
which ran in 1991, the Fermilab collaboration set limits on these decays as 
shown in 
Table~\ref{tab:limits}.  E799-II (part of KTeV) is expected to 
achieve sensitivities approaching SM predictions in some of these modes, and 
these sensitivities will be further improved by the subsequent KAMI (``Kaons 
at the Main Injector") program.

\begin{table}[htb]
\caption
{Limits on {\CP}-violating rare $K_L$ decays.}
\label{tab:limits}
%\vspace{5mm}
%\footnotesize
\begin{center}
\begin{tabular}{|l|c|c|c|}
\hline
Mode & E799-I limit & KTeV sens. & SM pred. \\
\hline\hline
$K_L\to\pi^0 e^+ e^-$ & $1.8\times10^{-9}$ & $7\times10^{-11}$ & $\sim10^{-11}$
\\
\hline
$K_L\to\pi^0 \mu^+ \mu^-$ & $5.1\times10^{-9}$ & few $10^{-11}$ &
$\sim10^{-11}$
\\
\hline
$K_L\to\pi^0 \nu{\overline\nu}$ & $5.8\times10^{-5}$ & $\sim10^{-8}$ &
$\sim\,$few $10^{-11}$ \\
\hline
\end{tabular}
\end{center}

\end{table}

\section{The Search for Direct {\CP} Violation in Hyperon Decay}

It has long been realized that hyperon decays could violate {\CP} 
symmetry~\cite{Pais}. 
Indirect {\CP} violation is not expected, since hyperon mixing would violate
conservation of baryon number.
Observables for direct {\CP} violation include 
decay-width differences of particle and antiparticle to 
{\CP}-conjugate final states and three asymmetries (described next) involving 
polarization.

In the decay of a polarized hyperon, the angular distribution of the daughter 
baryon in the rest frame of the parent is nonisotropic and is given by
\begin{equation}
\label{eq:ang-dist}
\frac{dN}{d\Omega} =\frac{1}{4\pi}(1+\alpha \vec{P}_p \cdot \hat{p}_d) = 
\frac{1}{4\pi}(1+\alpha P_p\cos{\theta})\,,
\end{equation}
where $\vec{P}_p$ is the polarization of the parent hyperon, $\hat{p}_d$ is
the direction of the daughter baryon in the rest frame of the parent,
and the parameter $\alpha$ is defined in Eq.~\ref{eq:SP}
below. Moreover, the daughter baryon is 
polarized, 
with polarization vector
\begin{equation}
\vec{P}_d = \frac{(\alpha+\vec{P}_p\cdot 
\hat{p}_d)\hat{p}_d+\beta(\vec{P}_p\times 
\hat{p}_d)+\gamma[\hat{p}_d\times(\vec{P}_p\times \hat{p}_d)]}
{1+\alpha \vec{P}_p\cdot \hat{p}_d}\,,
\end{equation}
where the Lee-Yang variables~\cite{Lee-Yang} $\alpha$,
$\beta$, and $\gamma$ are related to the $S$- and $P$-wave decay amplitudes:
\begin{equation}
\label{eq:SP}
\alpha = \frac{2\,{\rm Re}(S^*\!P)}{|S|^2+|P|^2}\,,~
\beta =\frac{2\,{\rm Im}(S^*\!P)}{|S|^2+|P|^2}\,,~
\gamma = \frac{|S|^2-|P|^2}{|S|^2+|P|^2}\,. 
\end{equation}
($\alpha$, $\beta$, and $\gamma$ are of course not all independent, being 
related by $\alpha^2+\beta^2+\gamma^2=1$.) Since under a {\CP} 
transformation $\alpha$ and $\beta$ change sign, in 
comparing the decays of a hyperon and its antiparticle we have the four
possibly-{\CP}-violating observables
\begin{equation}
\label{eq:alpha}
\Delta\equiv\frac{\Gamma-\overline{\Gamma}}{\Gamma+\overline{\Gamma}}\,,~ 
A\equiv\frac{\alpha+\overline{ \alpha}}{\alpha-\overline{ \alpha}}\,,~ 
B\equiv\frac{\beta+\overline{ \beta}}{\beta-\overline{ \beta}}\,,~  
B^\prime\equiv\frac{\beta+\overline{ \beta}}{\alpha-\overline{ \alpha}}\,,
\end{equation}
where $\Gamma\propto |S|^2+|P|^2$ is the partial decay width to a given
final state and
the overlined quantities pertain to antiparticles. As seen from 
Eq.~\ref{eq:SP}, nonzero values of $A$, $B$, and $B^\prime$
reflect interference 
between the $S$- and $P$-wave amplitudes.

As in the case of  the $K^0$, direct {\CP}-violating effects in hyperon decay 
arise in 
the SM via the interference of penguin and tree-level diagrams. Their size has 
been estimated using a variety of approaches~\cite{HyperCP}. $A$ is typically 
predicted to be of order $10^{-5}$ to $10^{-4}$ and is experimentally 
the most accessible; it can be measured by determining the daughter 
polarization in the decay of unpolarized parent hyperons. $B$ and $B^\prime$
are expected to be substantially larger than $A$ (and in the case of
$B^\prime$, independent of final-state phases) but require measurement  of 
both the parent and daughter polarizations. $\Delta$ is unobservably 
small. 

Although hyperon {\CP} asymmetries and $\epsilon^\prime$ arise from similar
quark diagrams, their SM phenomenologies are quite distinct.
$\epsilon^\prime$ arises from interference between $\Delta I=1/2$ and $\Delta 
I=3/2$ currents and is subject to the $m_t$-dependent cancellation mentioned
above.  On the other hand, $A$ is relatively insensitive to $m_t$,
with the central predicted value varying by only about $\pm15$\% for
$140<m_t<220\,$GeV in a typical calculation~\cite{He}.

Initial ideas for the measurement of $A$ centered on exclusive 
production of $\Lambda\overline{\Lambda}$ pairs in 
$\overline{p}p$ 
annihilation at low energy~\cite{Donoghue}. This technique has yielded the 
best result to date, $A=0.022 \pm0.019$~\cite{Barnes}. While experiments with 
substantially improved sensitivity have been proposed both for the LEAR 
storage ring at CERN~\cite{Hamann} and the $\overline{p}$ source at 
Fermilab~\cite{Hsueh}, none has yet been approved.\footnote{Sadly, with
LEAR now to be decommissioned, only one locus remains for such studies.} 

\subsection{The HyperCP experiment}

\begin{figure}[htb]
\vspace{0.in}
\centerline{\hspace{0.85in}\epsfysize = 6.5 in \epsffile {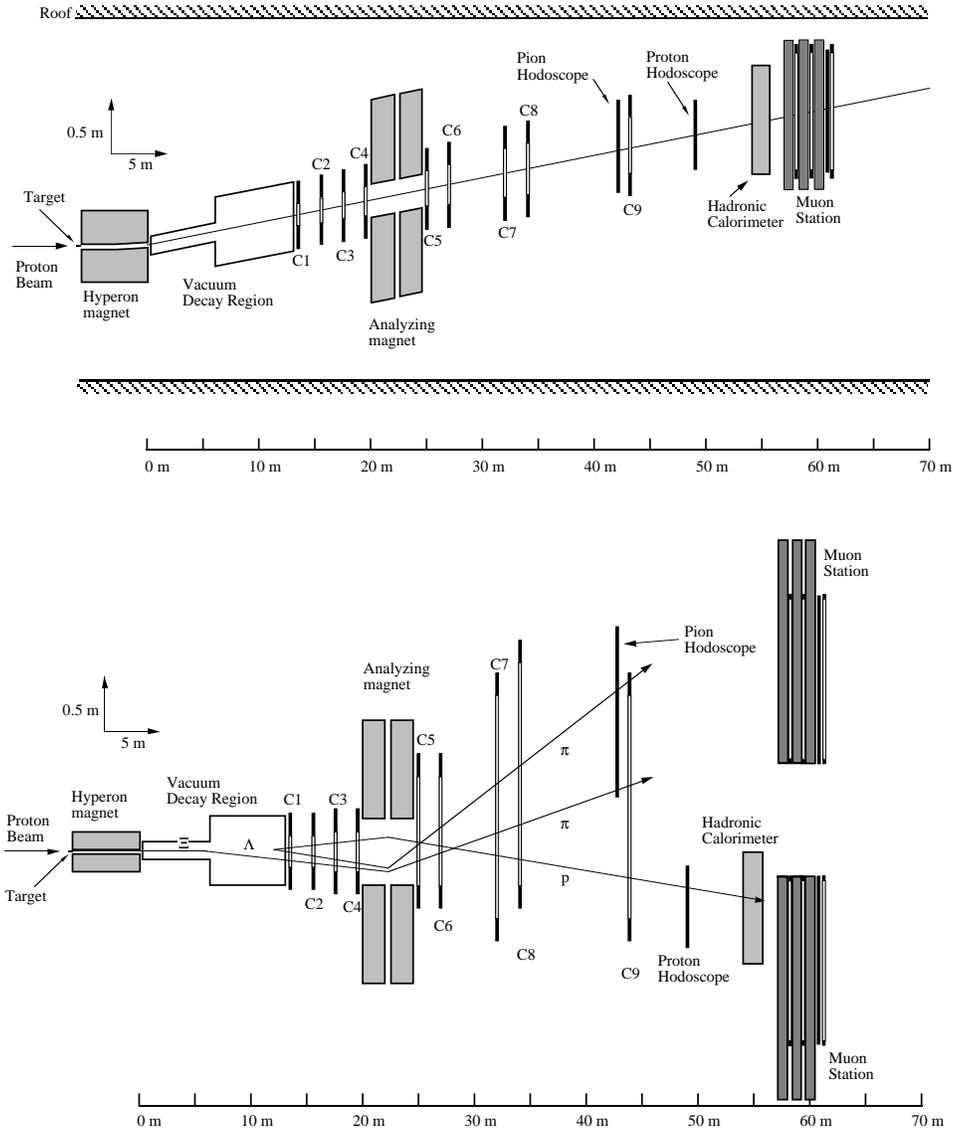}}
\vspace{-0.5in}
\caption {Elevation and plan views of HyperCP spectrometer.
\label{e871fig}}
\end{figure}

The HyperCP (E871) experiment~\cite{E871} (Fig.~\ref{e871fig}), 
now under construction by a 
Berkeley-Fermilab-Guanajuato-IIT-Michigan-S.~Alabama-Taiwan-Virginia
collaboration,\footnote{C. Ballagh, W. S. Choong, G. Gidal, P. Gu, K. B. Luk 
(Berkeley), 
T. Carter, C. James, J. Volk (Fermilab), 
J. Felix, G. Moreno, M. Sosa (Guanajuato),
R. A. Burnstein, A. Chakravorty, D. M. Kaplan, L. M.
Lederman, A. Ozturk, H. A. Rubin, D. Sowinski, C. White, S. White (IIT), 
H. R. Gustafson, M. Longo (Michigan),
K. Clark, M. Jenkins (S. Alabama), 
A. Chan, Y. C. Chen, K. C. Cheng,
C. Ho, M. Huang, P. K. Teng, C. Yu, Z. Yu (Taiwan), 
S. Conetti, C. Dukes, K. Nelson, D. Pocanic, D. Rajaram
(Virginia).
}
will measure the combined asymmetry in $\alpha$ in the decay sequence 
$\Xi^-\to\Lambda\pi^-$,
$\Lambda\to p\pi^-$. An intense unpolarized beam of $\Xi^-$ ($\overline 
{\Xi}^+$) hyperons will be produced at  $0^\circ$ by 800\,GeV protons striking 
a metal target, with the secondaries momentum-selected by  means of a curved 
magnetic channel set to 150\,GeV with 25\% FWHM momentum bite. Following a 
13\,m evacuated decay pipe the hyperon decay products will be detected in a 
high-rate magnetic spectrometer using MWPCs. 
(The needed rate capability is 
determined by the $\approx$40\,MHz of charged particles, dominantly pions and 
protons, emerging from the channel.) The polarization of the $\Lambda$s is 
measured by the slope of the $\cos{\theta}$ distribution of the protons in the 
$\Lambda$ 
rest frame (Eq.~\ref{eq:ang-dist}). From Eq.~\ref{eq:alpha} it is 
straightforward to show that the 
combined {\CP} asymmetry is well approximated by 
\begin{equation}
A_{\Xi\Lambda} \equiv \frac{\alpha_\Xi\alpha_\Lambda - \alpha_{\overline 
{\Xi}}\alpha_{\overline{\Lambda}}}{\alpha_\Xi\alpha_\Lambda + 
\alpha_{\overline{\Xi}}\alpha_{\overline{\Lambda}}} \cong A_\Xi+A_\Lambda\,.
\end{equation}
E871 aims to reconstruct $_\sim$\llap{$^>$}\,$3\times10^9$ each of $\Xi$ and 
${\overline \Xi}$ decays ($>$10$^3$ per second of beam), measuring 
$A_{\Xi\Lambda}$ to an uncertainty 
$_\sim$\llap{$^<$}0.8$\times10^{-4}$. 
As discussed further below, this sensitivity is in the range of 
asymmetry predicted by the SM, as well as by other possible models of 
{\CP} violation.
A previous fixed-target hyperon experiment, Fermilab E756 (which included some
of the HyperCP collaborators), is analyzing data on 
$A_{\Xi\Lambda}$ from the 1987/8 run with expected sensitivity 
$_\sim$\llap{$^<$}$10^{-2}$~\cite{Luk-private}.

The acquisition of so large a hyperon sample requires a highly capable
data acquisition system, designed for 100\,kHz trigger rate
and 20\,MB/s average data rate to tape. 
This high rate capability 
is driven by the need to use loose trigger requirements so as to
minimize any possible {\CP} bias~\cite{E871optrig}.
(KTeV is designing for
similar bandwidths, and as we will see, Charm2000 plans to go even further in 
data acquisition rate.)

Direct {\CP}-violating asymmetries are typically 
proportional to products of the weak-interaction and strong-interaction 
phase factors of the interfering decay amplitudes. The weak phases arise from 
short-distance physics, while the strong phases are due to final-state 
interactions. In the case of $A_{\Xi\Lambda}$, the strong phases in $\Lambda$ 
decay are directly measured in $\pi p$ scattering~\cite{Roper}, but those in 
$\Xi$ decay must be calculated theoretically. Older work relied on the 
calculations of Refs.~\cite{Nath} and \cite{Martin}, giving a phase difference 
of $16^\circ$, but a recent calculation using chiral perturbation theory gives 
$1.5^\circ$~\cite{Lu}, implying a $\Xi$ {\CP} asymmetry one order of magnitude 
smaller than previously thought. Thus in the SM $A_\Xi$ was formerly thought 
to exceed $A_\Lambda$, with predicted values in the range $\approx(0.1$ to 
$1)\times10^{-4}$ compared to a predicted range $\approx(0.1$ to 
$0.5)\times10^{-4}$ for $A_\Lambda$ (Table~\ref{tab:HyperCP}; \cf 
Ref.~\cite{HyperCP}). However, if the newer calculation is correct, 
$A_{\Lambda}$ is the larger contribution. At present it is not clear which (if 
either) calculation is correct~\cite{Suzuki}. A measurement of $\beta/\alpha$ 
using polarized $\Xi$s could help clarify the question. 

Hyperon {\CP} asymmetries have also been estimated in a variety of 
non-Standard models, and results are summarized in Table~\ref{tab:HyperCP}. In 
the Weinberg model and left-right-symmetric models with left-right mixing, 
$A_{\Xi\Lambda}$ can be substantially larger than in the SM, while in models 
in which {\CP} is violated due to FCNE it is essentially zero.

\begin{table}[htb]
\centering
\caption
{Hyperon {\CP}-asymmetry estimates.}
\label{tab:HyperCP}
%\vspace{5mm}
%\footnotesize
\begin{center}
\begin{tabular}{|l|c|c|c|}
\hline
& $A_\Xi$ & $A_\Lambda$ & \\
\raisebox{1.5ex}[0pt]{Model} & [$10^{-4}$] & [$10^{-4}$] & 
\raisebox{1.5ex}[0pt]{Ref.}\\
\hline
\hline
CKM & $-(0.1- 1)$ & $-(0.1- 0.5)$ &
\cite{He-Pakvasa} \\
Weinberg & $\approx\!-3.2$ & $\approx\!-0.25$ & \cite{Donoghue2} \\
Multi-Higgs (FCNE)  & $\approx$0 & $\approx$0 & 
\cite{He-Pakvasa} \\
LR (isoconjugate) & $\approx$0.25 & $\approx\!-0.11$ & \cite{Donoghue2} \\
LR (with mixing) & $<$1$\,^*$ & $<$7 & \cite{Chang}\\
\hline
\end{tabular}
\end{center}
\flushleft{$^*$using final-state phases of Ref.~\cite{Lu}}
\end{table}

\subsection{Sensitivity to charged-kaon direct {\CP} violation in HyperCP}

The HyperCP experiment also has the potential to observe direct {\CP} violation
in charged-kaon decay to $\pi^\pm\pi^\pm\pi^\mp$~\cite{E871}. 
The most accessible signal is the difference $\Delta g$
of the Dalitz-plot slope parameters for $K^+$ and $K^-$ decay that
measure the energy dependence of the odd-sign pion. SM predictions for $\Delta
g$ vary over a wide range, $\sim$10$^{-6}$ to $1.4\times10^{-3}$~\cite{KCP}.
The best previous measurement (from the Brookhaven AGS) 
gives $\Delta g=-0.0070\pm0.0053$~\cite{Kglimit}. 
HyperCP should amass a sample of $\approx$10$^9$ events in each mode, giving 
sensitivity of about $1\times
10^{-4}$. Other proposals are also extant at comparable
sensitivity~\cite{otherKCP}.

\section{The Search for {\CP} Violation in Charm Decay}

\begin{figure}[htb]
\vspace{0.1 in}
\centerline{\hspace{-.75in}\epsfysize=3 in\epsffile{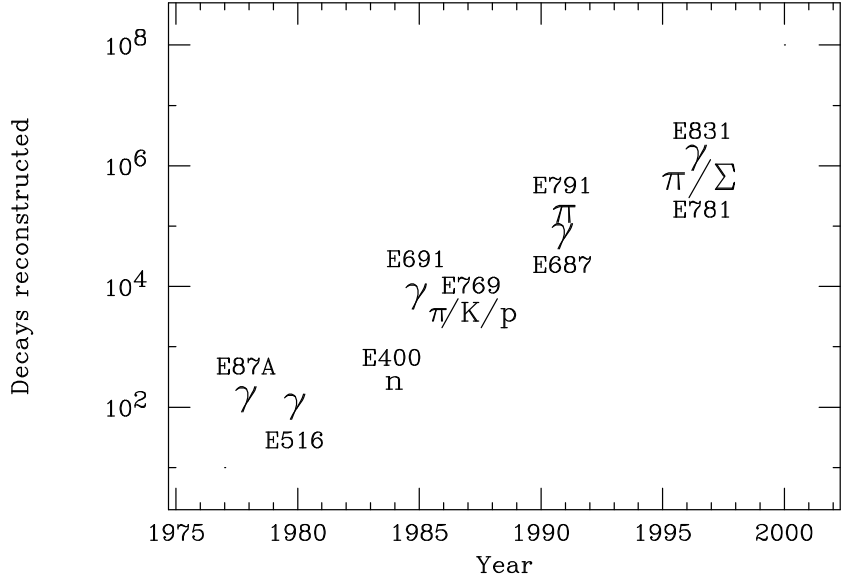}}
\caption{Yield of reconstructed charm \vs year of run
for those completed and approved Fermilab fixed-target charm experiments with
the highest statistics of their generation; symbols indicate type of beam
employed.\label{history}}
\end{figure}

Following the more-or-less simultaneous discovery of the charm quark in
fixed-target~\cite{Ting} and $e^+e^-$
collisions~\cite{Richter}, for many years 
experiments at $e^+e^-$ colliders dominated the study of charmed particles.
Starting in $\approx$1985, silicon vertex detectors made fixed-target 
experiments
competitive once again.
Although {\CP} asymmetries in charm are expected to be quite small,
exponential growth in the sensitivity of fixed-target charm experiments 
(Fig.~\ref{history}), as well as at
CLEO~\cite{Besson-Freyberger}, has  led to {\CP}-violation sensitivities that
are beginning to approach levels predicted in some 
extensions of the Standard Model.
As discussed below, the Charm2000 project at Fermilab may 
succeed in observing SM {\CP} violation.

\subsection{The Charm2000 project}

\begin{figure}[htb]
\centerline{\epsfysize = 1.89 in \epsffile {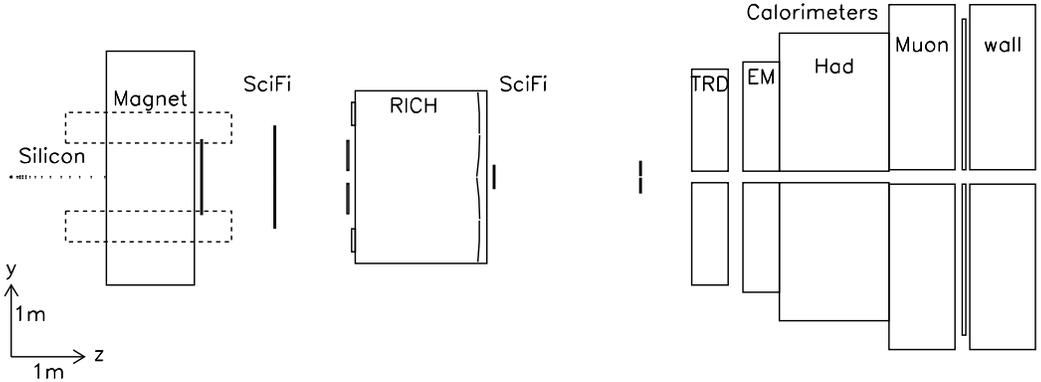}}
\caption {Charm2000 spectrometer concept (bend view).
\label{app865}}
\end{figure}

\begin{figure}[htb]
\vspace{0.1in}
\hspace{-.1in}\centerline{\epsfysize =1.5in \epsffile {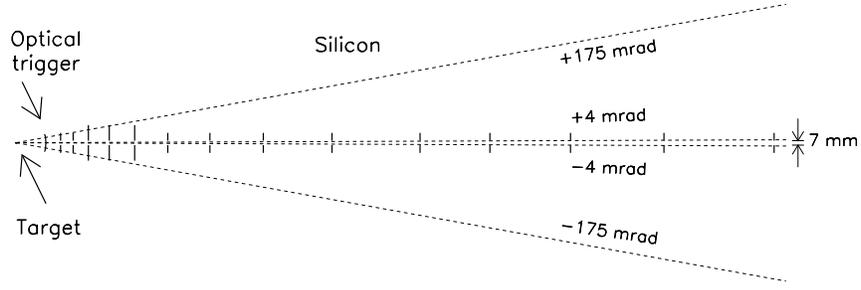}}
\caption {Detail of Charm2000 vertex region (showing optional optical
impact-parameter trigger).
\label{detail865}}
\end{figure}

Charm2000~\cite{Kaplan2000} is a Letter-of-Intent-in-progress for a new 
Fermilab experiment  to reconstruct $\approx$4$\times10^8$ charm decays
in the Year-$\approx$2000 
fixed-target run.
This sensitivity goal is $\approx$2000 times 
the largest extant charm sample, that of Fermilab E791.
The spectrometer (Figs.~\ref{app865}, \ref{detail865})
is planned to be compact and of moderate cost
(\eg substantially cheaper than HERA-$B$~\cite{HERA-B}), but
with large acceptance, good resolution, and high-rate 
tracking and particle identification. Tracking is done exclusively with
silicon or diamond~\cite{Tesarek} and 
scintillating-fiber~\cite{Ruchti} detectors, allowing operation at a 5\,MHz 
interaction rate. A
fast ring-imaging Cherenkov counter~\cite{Bari} 
provides hadron identification, and 
calorimeters (possibly augmented by a TRD) 
identify electrons and allow first-level triggering on transverse 
energy. 
Triggering efficiently on charm while maintaining high livetime and a
manageable data rate to tape ($_\sim$\llap{$^<$}100\,MB/s)
is a significant challenge,\footnote{While
HERA-$B$ is potentially competitive with Charm2000 as a
charm experiment, it lacks the capabilities to trigger efficiently on charm
and to acquire the needed large data sample, and it may have significantly
poorer vertex resolution as well.}
 requiring
hardware decay-vertex triggers~\cite{triggers}; first-level ``optical" triggers
may play a significant role~\cite{optrig,mul-jump}. (More detailed discussions
of the Charm2000 spectrometer and physics goals
may be found in~\cite{Kaplan2000}.)

\subsection{Direct charm {\CP} violation}

The Standard Model predicts direct {\CP} violation in 
singly Cabibbo-suppressed decays (SCSD) of charm  at the $\sim$10$^{-3}$ 
level~\cite{CharmCP,Burdman}.
{\CP} violation  in
Cabibbo-favored (CFD) or doubly Cabibbo-suppressed (DCSD) modes would 
be a clear signature for new physics~\cite{Burdman,Bigi94}. 
Asymmetries in all 
three  categories could reach $\sim$10$^{-2}$ in such scenarios as 
non-minimal supersymmetry~\cite{Bigi94} and in left-right-symmetric 
models~\cite{left-right,Pakvasa2000}. There are also expected SM asymmetries of
$\approx\!3.3\times10^{-3}$ ($=\!\!2\,$Re$(\epsilon_K)$) due to $K^0$ mixing in
such modes as $D^+\to K_S\pi^+$ and $K_S \ell\nu$~\cite{Xing}, which should be
observed in Charm2000 or even in predecessor experiments~\cite{CHEOPS}. 
While $K^0$-induced {\CP} asymmetries might
teach us little we do not already know,
they will at least constitute a calibration
for experimental systematics at the $10^{-3}$ level.
However, Bigi has pointed out that a small new-physics contribution to the DCSD
rate could amplify these asymmetries to $\cal{O}$$(10^{-2})$~\cite{Bigi94}.

Experimental limits at the 10\% level have been set in SCSD modes; at present 
the most sensitive 
come from the photoproduction experiment Fermilab E687~\cite{Frabetti} and 
from CLEO~\cite{Bartelt}. E687 has studied $D^0\to K^+K^-$ and $D^+\to 
K^-K^+\pi^+$, $\overline {K^{*0}}K^+$, 
and $\phi\pi^+$ 
as indicated in Table~\ref{tab:charmCP}.\footnote{Charge-conjugate states
are generally included even when not 
stated.} CLEO has studied $D^0$ decays to $K^-\pi^+$ and to
the {\CP} eigenstates 
$K^+K^-$, 
$K_S\phi$, 
and $K_S\pi^0$.

\begin{table}[htb]
\caption{Limits on direct {\CP} violation in $D$ decay. \label{tab:charmCP}}
%\vspace{0.1in}
%\footnotesize
\begin{center}
\begin{tabular}{|l|l|l|}
\hline
& & Charm2000 \\
\raisebox{1.5ex}[0pt]{Mode} & \raisebox{1.5ex}[0pt]{Limit$^*$} & Reach$^*$
\\
\hline
\hline \multicolumn {3}{|l|}{Cabibbo-favored} \\ \hline
~$D^0\to K^- \pi^+$ & -0.009$<A<$0.027~\cite{Bartelt} &  \\
~$D^0\to K^- \pi^+\pi^-\pi^+$ & & few$\times10^{-4}$  \\
\hline \multicolumn {3}{|l|}{Singly Cabibbo-suppressed} \\ \hline
~$D^0\to K^- K^+$ & -0.11$<A<$0.16~\cite{Frabetti} & $10^{-3}$ \\
 & -0.028$<A<$0.166~\cite{Bartelt} & \\
~$D^+\to K^- K^+\pi^+$ & -0.14$<A<$0.081~\cite{Frabetti} & $10^{-3}$ \\
~$D^+\to \overline {K^{*0}}K^+$ & -0.33$<A<$0.094~\cite{Frabetti} &
$10^{-3}$ \\
~$D^+\to \phi\pi^+$ & -0.075$<A<$0.21~\cite{Frabetti} & $10^{-3}$ \\
~$D^+\to K_S\pi^+$ & & few$\times10^{-4}$ \\
\hline \multicolumn {3}{|l|}{Doubly Cabibbo-suppressed} \\ \hline
~$D^0\to K^+ \pi^-$ & & $10^{-3}-10^{-2}$ \\
~$D^+\to K^+ \pi^+ \pi^-$ & & few$\,\times10^{-3}$ \\
\hline
\end{tabular}
\end{center}
$^*$at 90\% confidence level
\end{table}

The signal for direct {\CP} violation is an absolute rate difference between 
decays of particle and antiparticle to charge-conjugate final states $f$ and 
${\bar f}$:
\begin{equation}
A=\frac{\Gamma(D\to f)-\Gamma({\overline D}\to{\bar f})}
{\Gamma(D\to f)+\Gamma({\overline D}\to{\bar f})}\,.
\end{equation}
Since in photoproduction $D$ and ${\overline  D}$ are not produced equally, in
the E687 analysis the signal is normalized relative to a CFD mode:
\begin{equation}
A=\frac{\eta(D\to f)-\eta({\overline  D}\to{\bar f})}
{\eta(D\to f)+\eta({\overline  D}\to{\bar f})}\,,
\end{equation}
where
\begin{equation}
\eta(D^0)=\frac{N(D^0\to K^+K^-)}{N(D^0\to K^-\pi^+)}\,,
\end{equation}
for the $D^+$ modes the normalization mode is $D^+\to K^-\pi^+\pi^+$, etc.
 (Thus 
a {\CP} asymmetry from new physics in 
the CFD normalization mode could in principal mask a signal in an SCSD mode.) 
A further complication 
is that to distinguish \eg $D^0\to K^+K^-$ from $\overline {D^0}\to K^+K^-$,
$D^*$ tagging (via the charge of the pion from $D^{*+}\to D^0\pi^+$)
must be employed; of course, no tagging is needed for charged-$D$ 
decays. Typical E687 event yields are $\approx$10$^2$ in signal modes and 
$\sim$10$^3$ 
in normalization modes.

 One can extrapolate from the sensitivity achieved in E687 to that 
expected in Charm2000. E687 observed $4287\pm78$ ($4666\pm81$) events in the 
normalization mode $D^+\to K^-\pi^+\pi^+$ ($D^-\to K^+\pi^-\pi^-$). As an 
intermediate step in the extrapolation I use the event yield in E791, since 
that hadroproduction experiment is more similar to Charm2000 than is E687. 
Using relatively tight vertex cuts, E791 observed $37006\pm 204$ events in 
$D^\pm\to K\pi\pi$~\cite{Aitala}, and Charm2000 should increase this number by 
a factor $\approx$2000.
 Thus relative to E687, the statistical 
uncertainty on $A$ should be reduced by $\approx\!\sqrt{8000}$, implying 
sensitivities in SCSD modes of $10^{-3}$ at 90\% confidence. While 
the 
ratiometric nature of the measurement reduces biases, at the $10^{-3}$ level 
these will need to be studied carefully. 

Since one CFD mode must be used for normalization, the search for direct
{\em CP} violation in CFD modes is actually a search for 
{\em differences} among various modes. Given the differing final-state
interactions~\cite{Buccella}, 
if new physics causes {\em CP} violation in CFD modes,
such {\em CP}-asymmetry differences are not unlikely.
The estimated event yields in Charm2000 imply
{\em CP} sensitivity at the few\,$\times10^{-4}$ level
for $D^0\to K^- \pi^+\pi^-\pi^+$, normalized to the
production asymmetry observed in $D^0\to K^- \pi^+$.

For DCSD modes, I extrapolate from E791's observation of $D^+\to 
K^+\pi^+\pi^-$ at $4.2\sigma$ based on 40\% of their data 
sample~\cite{Purohit-Weiner}. The statistical significance in Charm2000 should 
be $\approx\!\sqrt{2000/0.4}$ better, implying few$\times10^{-3}$ sensitivity 
for {\CP} asymmetries. For $D^0\to K^+\pi^-$,
CLEO's observation~\cite{Cinabro} of $B(D^0\to K^+\pi^-)/B(D^0\to
K^-\pi^+)\approx0.8$\% suggests $\approx\!10^5$ $D^*$-tagged DCSD $K\pi$
events in
Charm2000, giving few$\times10^{-3}$ {\CP} sensitivity. However, the need
for greater background suppression for DCSD compared to CFD events
is likely to reduce sensitivity. For example, preliminary E791 results show
a $\approx$2$\sigma$ signal in $D^0\to K^+\pi^-$~\cite{Purohit95}, 
implying $\sim$10$^{-2}$ sensitivity in Charm2000. 

Table~\ref{tab:charmCP} summarizes the above estimates of Charm2000 {\em 
CP}-violation sensitivity.
These extrapolations are conservative insofar as they
ignore expected  improvements in
vertex resolution and particle identification. Simulations  are
underway to assess these effects. 

As in the kaon and hyperon cases, SM predictions for direct charm {\CP}
violation are rather uncertain, requiring assumptions for final-state phase
shifts and CKM matrix elements~\cite{Burdman,Bigi94}.
However, given the order of magnitude
expected in charm decay, the Charm2000 experiment could make the first
observation of direct {\CP} violation outside the strange sector, or indeed 
the first observation anywhere if (as may well be the case~\cite{K0CP,Lu})
signals prove too small for detection in KTeV, NA48, and HyperCP.

\subsection{Indirect charm {\CP} violation}

Indirect {\CP} violation of course requires mixing,
but experimentally the $D^0$ mixing rate is known to be small 
($r_{\rm mix}<0.37$\%)~\cite{PDG,E691}. 
For small mixing, the mixing rate is given to good approximation
by~\cite{Burdman}
\begin{equation}
r_{\rm mix}\approx \frac{1}{2}\bigg [\bigg (\frac{\Delta M_D}{\Gamma_D}
\bigg )^2 + \bigg (\frac{\Delta\Gamma_D}{2\Gamma_D}\bigg )^2\bigg ]\,.
\end{equation}
In the SM, the $\Delta M$ and $\Delta\Gamma$
contributions are expected to be of the same order of
magnitude and are estimated~\cite{Burdman,mixing} to give
$r_{\rm mix}<10^{-8}$;\footnote{Earlier
estimates~\cite{Wolfenstein_mixing}
that long-distance effects can give $|\Delta M_D/\Gamma_D |\sim 10^{-2}$
are claimed to have been disproved~\cite{Burdman,Hall-Weinberg}, though there
 remain skeptics~\cite{Bigi94,Wolfenstein}.} 
any indirect {\CP}-violating asymmetries are
expected to  be less than $10^{-4}$~\cite{Bigi94}.
However, possible mixing signals at the $\approx$1\% level
have been reported~\cite{Cinabro,MkII-mixing}, and
a variety of non-Standard models
can accommodate mixing up to the current experimental limit, including
multi-Higgs 
models~[31, 89-93]
and those with
supersymmetry~\cite{Bigietal,Datta,Nir}, technicolor~\cite{Techni},
leptoquarks~\cite{Lepto}, left-right symmetry~\cite{seesawLR},
or a fourth generation~\cite{Pakvasa2000,Babu}.

$D^0$ mixing phenomenology
is complicated by the possibility of DCSD leading to the
same final states. For example in the $K\pi$ mode the rate of wrong-sign $D^0$
decay is given by~[99-101]
\begin{equation}
\Gamma(D^0(t)\to K^+\pi^-) = |B|^2|\frac{q}{p}|^2
\frac{e^{-\Gamma t}}{4}
\{4 |\lambda|^2
+ (\Delta M^2 + \frac{\Delta\Gamma^2}{4})t^2 + 
2{\rm Re}(\lambda)\Delta\Gamma t + 4 {\rm Im}(\lambda)\Delta Mt\}\,,
\label{eq:mixing}
\end{equation}
where the first term is the DCSD contribution, the second mixing, and the third
and fourth the interference between DCSD and mixing.
Given the E691 mixing limit~\cite{E691}, the observed signals 
presumably
represent enhanced DCSD effects. If a significant portion of this rate
is mixing, new physics must be responsible~\cite{Burdman,Wolfenstein}, and
indirect {\em CP} violation at the $_\sim$\llap{$^<$}1\%
level is then possible~\cite{Bigi-Sanda,Bigi2000,Bigi94,Wolfenstein}.
Several authors have suggested that the {\em CP}-violating signal, which
arises from the interference term of Eq.~\ref{eq:mixing}, may be
easier to detect than the mixing
itself~[100-102, 88]. In particular, Browder and
Pakvasa~\cite{Browder} point out that in the difference
$\Gamma(D^0\to K^+\pi^-)-\Gamma(\overline{D^0} \to K^-\pi^+)$, the DCSD
and mixing components cancel, leaving only the fourth term of
Eq.~\ref{eq:mixing}. Thus if indirect {\em CP} violation is
appreciable, this is a particularly clear way to isolate the
interference term.

\label{sec:mixing}

Whether or not it violates {\em CP},
$ D^0\overline {D^0} $ mixing may be one of the more promising places to 
look for low-energy manifestations of physics beyond the Standard Model.
An interesting example  is the multiple-Higgs-doublet model 
lately expounded by Hall and Weinberg~\cite{Hall-Weinberg}, in  which
$|\Delta M_D|$ can be as large as $10^{-4}$\,eV, approaching the current 
experimental limit. In this model all {\em CP} violation arises from 
flavor-changing neutral-Higgs exchange and 
is intrinsically of order 10$^{-3}$, too small to be 
observed in the beauty sector and (except through mixing) in the kaon sector,
but (as mentioned above)
possibly observable in charm -- another example of the importance of
exploring rare phenomena in {\em
all} quark sectors. 
Multiple-Higgs models are one of the simplest extensions of
the SM~\cite{Pakvasa2000,Winstein-Wolfenstein,Wu}, and advantage should be
taken of all opportunities to test them.

Clarification of the $D^0$ mixing puzzle can be expected from coming
experiments as well as (if approved) Charm2000. If mixing is large 
and violates {\CP} as just discussed, indirect {\CP} violation
should be detectable in Charm2000.

\section{Conclusions}

{\CP} violation has fascinated physicists since its discovery.
It has the potential to give us unique information about the
physics underlying the Standard Model. In the down-quark sector the phenomenon 
may be dominated by Standard-Model effects arising from the large mass of the 
top quark, but in the up-quark sector the $b$ quark contributes little, 
creating a low-background window for new physics. 
If new physics and the CKM phase are both significant sources of {\CP} 
violation, then coming beauty studies will reveal deviations of the 
CKM-matrix ``unitarity triangle" from expectations~\cite{Rosner}. But if either 
contribution is small, these studies might tell us little new: in the one case 
the unitarity triangle will confirm the CKM model, while in the other, beauty 
decays might not violate {\CP} at an observable level. New physics might still 
be revealed in hyperon or charm studies.
A program investigating all possible quark sectors is thus prudent. The 
Fermilab fixed-target program can make a strong contribution to such a program.


\begin{thebibliography}{999}

\bibitem{Cronin-Fitch}
J. H. Christenson, J. W. Cronin, V. L. Fitch, and R. Tourlay, Phys.\ Rev.\ 
Lett.\ {\bf 13}, 138 (1964).

\bibitem{Sakharov}
A. D. Sakharov, Pis'ma Zh.\  Eksp.\ Teor.\ Fiz.\ {\bf 5}, 32 (1967) [JETP 
Lett.\ {\bf 5}, 24 (1967)].

\bibitem{KM}
M. Kobayashi and T. Maskawa, Prog.\ Theor.\ Phys.\ {\bf 49}, 652 (1973).

\bibitem{Heinrich}
J. Heinrich, E. A. Paschos, J. M. Schwarz, and Y. L. Wu,
 Phys.\ Lett.\ B {\bf 279}, 140 (1992).

\bibitem{K0CP}
For recent calculations, see 
M. Ciuchini, E. Franco, G. Martinelli, and L. Reina,
Phys.\ Lett.\ B {\bf 
301}, 263 (1993);\hfill\break
A. J. Buras, M. Jamin, and M. E. Lautenbacher, Nucl.\ Phys.\ 
B (1993);\hfill\break
M. Ciuchini {\it et al.}, 
 ``Estimates of $\epsilon^\prime/\epsilon$,"
ROME-1091-1995, hep-ph/9503277, to appear in the  2nd Daphne Physics Handbook
(1995).

\bibitem{KCP}
G. D'Ambrosio, G. Isidori, and N. Paver, Phys.\ Lett.\  B {\bf 273}, 497 
(1991);\hfill\break 
H. Y. Cheng, Phys.\ Rev.\ D {\bf 44}, 919 (1991);\hfill\break
A. A. Bel'kov {\it et al.}, Phys.\ Lett.\  B {\bf 300}, 283 (1993).

\bibitem{HyperCP}
For 
recent summaries see 
J. F. Donoghue, B. R. Holstein, and 
G. Valencia, Int.\ J.\ Mod.\ Phys.\ A {\bf 2}, 319 (1987);\hfill\break
X. G. He and S. Pakvasa, to appear in {\sl Proc.\ DPF
'94}, Albuquerque, NM, Aug.\ 1--8, 1994; and\hfill\break 
N. G. Deshpande, X. G. He, and S. 
Pakvasa,  Phys.\ Lett.\ B {\bf 326}, 307 (1994).

\bibitem{CharmCP}
M. Golden and B. Grinstein, Phys.\ Lett.\ B {\bf 222}, 501 (1989);\hfill\break 
F. Buccella {\it et al.}, Phys.\ Lett.\ B {\bf 302}, 319 (1993) and
Phys.\ Rev.\ D {\bf 51}, 3478 (1995).

\bibitem{Rosner}
J. L. Rosner, ``Present and Future Aspects of {\CP} Violation," 
EFI 95-36, hep-ph/9506364, to appear in 
{\sl Proc.\ LAFEX International School on High Energy Physics (LISHEP95)},
 Rio de Janeiro, Brazil, Feb.\ 
7--22, 1995.

\bibitem{Cabibbo}
N. Cabibbo, Phys.\ Rev.\ Lett.\ {\bf 10}, 531 (1963).

\bibitem{Nozaki}
T. Nozaki, ``The KEK B Factory and the BELLE Detector,"
KEK-PREPRINT-95-168, Nov.\ 1995, to appear in
{\sl Proc.\ Conf.\  on Production and Decay of
Hyperons, Charm and Beauty Hadrons},
Strasbourg, France, 5--8 Sept.\ 1995.

\bibitem{PEP-II}
G. Wormser,  Nucl.\ Instr.\ Meth.\ {\bf A351}, 54 (1994). 

\bibitem {HERA-B}
T. Lohse {\it et al.}, ``HERA-$B$:
An Experiment to Study {\CP} Violation in the $B$
System Using an Internal Target at the HERA Proton Ring," Proposal to DESY,
DESY-PRC 94/02, May 1994.

\bibitem{DeJongh}
F. DeJongh,  ``$B$ Physics with the CDF Run II Upgrade,"
FERMILAB-CONF-95-408-E, hep-ex/9512008, Dec.\ 1995, to appear in
{\sl Proc.\ Conf.\  on Production and Decay of
Hyperons, Charm and Beauty Hadrons},
Strasbourg, France, 5--8 Sept.\ 1995.

\bibitem{Berkelman}
K. Berkelman, ``{\em CP} Violation at a Symmetric $e^+ e^-$ Collider,"
CLNS-95-1322, May 1995,
to appear in {\sl Proc.\ Lafex International School on High
Energy Physics (LISHEP95)},
Rio de Janeiro, Feb.\ 7-22, 1995. 

\bibitem{Winstein-Wolfenstein}
B. Winstein and L. Wolfenstein, Rev.\ Mod.\ Phys.\ {\bf 65}, 1113 (1993).

\bibitem{E871}
J. Antos {\it et al.}, Fermilab Proposal 871 (revised), Mar.\ 26, 
1994; see also \hfill\break 
E. C. Dukes, ``A New Fermilab Experiment to Search for Direct {\CP}
 Violation in 
Hyperon Decays," to appear in {\sl Proc.\ 11th Int.\
Symp.\ on High Energy Spin Physics},  Bloomington,
IN, Sept.\ 15--22, 1994.

\bibitem{superweak}
L. Wolfenstein, Phys.\ Rev.\ Lett.\ {\bf 13}, 562 (1964).

\bibitem{Ellis}
A. I. Vainshtein, V. I. Zakharov, and M. A. Shifman, LETP Lett.\ {\bf 22}, 55
(1975);\hfill\break
V. I. Zakharov, M. A. Shifman, and A. I. Vainshtein, Nucl.\ Phys.\ {\bf B120},
316 (1977);\hfill\break
J. Ellis, M. K. Gaillard, D. V. Nanopoulos, and S. Rudaz, Nucl.\ Phys.\
{\bf B131}, 285 (1977).

\bibitem{Flynn_Buchalla}
J. M. Flynn and L. Randall, Phys.\ Lett.\ B {\bf 224}, 221 (1989);\hfill\break
G. Buchalla, A. J. Buras, and M. Harlander, Nucl.\ Phys.\ {\bf B337}, 313 
(1990).

\bibitem{Paschos}
E. A. Paschos and Y. L. Wu, Mod.\ Phys.\ Lett.\ {\bf A6}, 93 (1991).

\bibitem{mtop}
F. Abe {\it et al.}, Phys.\ Rev.\ Lett.\ {\bf 74}, 2626 (1995);\hfill\break 
S. Abachi {\it et al.}, {\it ibid.}, p.\ 2632.

\bibitem{Paschos2} E. A. Paschos, private communication.

\bibitem{Donoghue}
See for example Donoghue {\it et al.}, {\it op. cit.} 
Ref.~\protect\cite{HyperCP}.

\bibitem{Lee}
T. D. Lee, Phys.\ Rev.\ D {\bf 8}, 1226 (1973) and
Phys.\ Rep.\ {\bf 9C}, 143
(1974).

\bibitem{Weinberg}
S. Weinberg, Phys.\ Rev.\ Lett.\ {\bf 37}, 657 (1976).

\bibitem{Bigi-Sanda}
I. I. Bigi and A. Sanda, in {\bf CP Violation}, C. Jarlskog, {\it ed.}, World
Sceintific, Singapore (1989), p.\ 362.

\bibitem{PDG}
L. Montanet {\it et al.} (Particle Data Group), Phys.\ Rev.\ D {\bf 50}, 1173 
(1994).

\bibitem{tree}
S. Pakvasa and H. Sugawara, Phys.\ Lett.\ {\bf
73B}, 61 (1978);\hfill\break 
S. Pakvasa {\it et al.}, Phys.\ Rev.\ D {\bf 25}, 1895
(1982);\hfill\break 
T. P. Cheng and M. Sher, Phys.\ Rev.\ D {\bf 35}, 3484 (1987);\hfill\break
M. Shin, M. Bander, and D. Silverman, in {\sl Proc. Tau-Charm Factory
Workshop}, Stanford Linear Accelerator Center,
Stanford, CA, May 23--27, 1989, SLAC-Report-343, p.\ 686.

\bibitem{Weinberg-Glashow}
S. L. Glashow and S. Weinberg, Phys.\ Rev.\ D {\bf 15}, 1958 (1977).

\bibitem{Hall-Weinberg}
L. Hall and S. Weinberg, Phys.\ Rev.\ D {\bf 48}, R979 (1993).

\bibitem{Wu}
Y. L. Wu and L. Wolfenstein, Phys.\ Rev.\ Lett.\ {\bf 73}, 1762 (1994).

\bibitem{Mohapatra}
R. N. Mohapatra, in {\bf {\CP} Violation}, C. Jarlskog, {\it ed.}, World
Sceintific, Singapore (1989), p.\ 384.

\bibitem{Mohapatra-Pati}
R. N. Mohapatra and J. C. Pati, Phys.\ Rev.\ D {\bf 11}, 566 (1975).

\bibitem{Gibbons}
L. K. Gibbons {\it et al.}, Phys.\ Rev.\ Lett.\ {\bf 70}, 1203 (1993).

\bibitem{Barr}
G. D. Barr {\it et al.}, Phys.\ Lett.\ B {\bf 317}, 233 (1993).

\bibitem{Woods}
M. Woods  {\it et al.}, Phys.\ Rev.\ Lett.\ {\bf 60}, 1695 (1988).

\bibitem{KTeV}
K. Arisaka {\it et al.}, Fermilab Proposal 832 (1990).

\bibitem{NA48}
G. D. Barr {\it et al.}, Proposal for NA48, CERN/SPSC/90-22 (1990).

\bibitem{Barmin}
V. Barmin {\it et al.}, Nucl.\ Phys.\ {\bf B247}, 293 (1984) and
{\bf B254},
747(E) (1984);\hfill\break 
T. Nakada, in {\bf Lepton and Photon Interactions}, P.
Drell and D. Rubin, {\it eds.}, AIP Conf.\ Proc.\ No.\ 302, American Institute
of Physics, New York (1994).

\bibitem{Schwingenheuer}
B. Schwingenheuer {\it et al.}, Phys.\ Rev.\ Lett.\ {\bf 74}, 4376 (1995).

\bibitem{GibbonsCPT}
L. K. Gibbons {\it et al.}, Phys.\ Rev.\ Lett.\ {\bf 70}, 1199 (1993).

\bibitem{Pais}
A. Pais, Phys.\ Rev.\ Lett.\ {\bf 3}, 242 (1959);\hfill\break 
O. E. Overseth and S. Pakvasa, Phys.\ Rev.\ {\bf 184}, 1663 (1969).

\bibitem{Lee-Yang}
T. D. Lee and C. N. Yang, Phys.\ Rev.\ {\bf 108}, 1645 (1957).

\bibitem{He}
X. G. He, H. Steger, and G. Valencia, Phys.\ Lett.\ B {\bf 272}, 411 (1991).

\bibitem{Barnes}
P. D. Barnes  {\it et al.}, in preparation.

\bibitem{Hamann}
N. Hamann {\it et al.}, CERN/SPSLC 92-19 (1992).

\bibitem{Hsueh}
S. Y. Hsueh and P. Rapidis, ``Search for Direct {\CP} Violation in 
$\overline{p} +p\to\overline{\Lambda}+ \Lambda\to\overline{p}\pi^+ + p\pi^-$,"
Proposal to Fermilab,  Jan.\ 2, 1992.

\bibitem{Luk-private}
K. B. Luk, private communication.

\bibitem{E871optrig}
The possibility of enriching the sample during a portion of the run
using an optical impact-parameter trigger device is however under investigation;
see 
M. Atac {\it et al.}, ``The Development of the Optical Discriminator," to
appear in {\sl Proc. 7th Vienna Wire Chamber Conf.}, Vienna, Austria,
13--17 Feb.\ 1995;\hfill\break
G. Charpak {\it et al.}, ``The Optical Trigger for the E871 Experiment,"
RD30 Note, March 16, 1995.

\bibitem{Roper}
L. Roper {\it et al.}, Phys.\ Rev.\ {\bf 138}, 190 (1965).

\bibitem{Nath}
R. Nath and B. Kumar, Nuov.\ Cim.\ {\bf 36}, 669 (1965).

\bibitem{Martin}
B. Martin, Phys.\ Rev.\ {\bf 138}, 1136 (1965).

\bibitem{Lu}
M. Lu, M. Savage, and M. Wise, Phys.\ Lett.\ B {\bf 337}, 133 (1994). 

\bibitem{Suzuki}
M. Suzuki, private communication to K. B. Luk.

\bibitem{Kglimit}
W. T. Ford {\it et al.}, Phys.\ Rev.\ Lett.\ {\bf 25}, 1370 (1970).

\bibitem{He-Pakvasa}
X. G. He and S. Pakvasa, {\it op cit.} Ref.~\protect\cite{HyperCP}.

\bibitem{Donoghue2}
J. F. Donoghue, X. G. He, and S. Pakvasa, Phys.\ Rev.\ D {\bf 34}, 833 (1986).

\bibitem{Chang}
D. Chang, X. G. He, and S. Pakvasa, Phys.\ Rev.\ Lett.\ {\bf 74}, 3927 (1995).

\bibitem{otherKCP}
See G. D. Barr {\it et al.}, Ref.~\protect\cite{NA48};\hfill\break 
M. Zeller
{\it et al.}, BNL Proposal 865 (1991);\hfill\break 
P. Franzini, INFN preprint LNF-92/024
(P) (1992).

\bibitem{Ting}
J. J. Aubert {\it et al.}, 
Phys.\ Rev.\ Lett.\ {\bf 33}, 1404 (1974).

\bibitem{Richter}
J. E. Augustin {\it et al.}, Phys.\ Rev.\ Lett.\ {\bf 33}, 1406 (1974).

\bibitem{Besson-Freyberger}
D. Z. Besson and A. P. Freyberger,
in {\bf The Future of High-Sensitivity Charm
Experiments}, {\sl Proc.\ CHARM2000 Workshop},
Fermilab, June 7--9, 1994,
D. M. Kaplan and S. Kwan, {\it eds.}, FERMILAB-Conf-94/190 (1994), p.\ 35.

\bibitem{Kaplan2000}
D. M. Kaplan, in {\bf The Future of
High-Sensitivity Charm Experiments}, {\it op cit.}, p.\ 229;\hfill\break
D. M. Kaplan and V. Papavassiliou, ``An Ultrahigh-Statistics Charm Experiment 
for the Year
$\sim$2000," 
IIT-HEP-95/2, hep-ex/9505002,
to appear in 
{\sl Proc.\ LAFEX International School High Energy Physics (LISHEP95)},
Rio de Janeiro, Brazil,
Feb.\ 7--22, 1995;\hfill\break
D. M. Kaplan, in {\bf Workshop on the Tau/Charm Factory},
Argonne National Laboratory, June 21--23, 1995, J. Repond, {\it ed.},
AIP Conf.\ Proc.\ No.\ 349, American Institute of Physics (1996), p.\ 
425;\hfill\break
D. M. Kaplan, ``Charm2000: An Ultrahigh-Statistics Charm Experiment for the
Turn of the Millennium," IIT-HEP-95/7, hep-ex/9512002,
to appear in {\sl Proc.\ Conf.\  on Production and Decay of
Hyperons, Charm and Beauty Hadrons},
Strasbourg, France, 5--8 Sept.\ 1995.

\bibitem{Tesarek}
R. Tesarek, in {\bf The
Future of High-Sensitivity Charm Experiments}, {\it op cit.}, p.\ 163.

\bibitem{Ruchti}
R. Ruchti, in {\bf The
Future of High-Sensitivity Charm Experiments}, {\it op cit.}, p.\
173;\hfill\break
D. Adams {\it et al.}, in {\bf 
4th Int.\ Conf.\ on Advanced Technology and Particle Physics}, Como,
Italy, 3--7 Oct.\ 1994, E. Borchi, S. Majewski, J.
Huston, A. Penzo, P.G. Rancoita, {\it eds.},
Nucl.\ Phys.\ B Proc.\ Suppl.\ {\bf 44}, Nov.\ 1995, p.\ 332.

\bibitem{Bari}
See e.g.\ D. M. Kaplan {\it et al.},
Nucl.\ Instr.\  Meth.\ {\bf A343}, 316 (1994);\hfill\break
D. F. Anderson, S. Kwan, and V. Peskov, {\it ibid.}, p. 109;\hfill\break
N. S. Lockyer {\it et al.}, {\it ibid.} {\bf A332}, 142 (1993);\hfill\break
T. Lohse {\it et al.}, {\it op cit.} Ref.~\protect\cite{HERA-B}.

\bibitem{triggers}
Further discussion of triggers for
heavy-quark experiments may be found in
D. Christian, in {\bf The Future of
High-Sensitivity Charm Experiments}, {\it op cit.}, p.\ 221, and D. Barberis,
{\it ibid.} p.\ 213.

\bibitem{optrig}
G. Charpak, Y. Giomataris, and L. M. Lederman, Nucl.\ Instr.\ Meth.\ {\bf
A306}, 439 (1991);\hfill\break 
D.~M.~Kaplan {\it et al.}, {\it ibid.} {\bf A330}, 33
(1993);\hfill\break 
G. Charpak {\it et al.}, {\it ibid.} {\bf A332}, 91 (1993).

\bibitem{mul-jump}
A. M. Halling and S. Kwan, Nucl.\ Instr.\ Meth.\ {\bf A333}, 324 (1993).

\bibitem{Burdman}
G. Burdman, in {\bf The Future of High-Sensitivity Charm
Experiments}, {\it op cit.}, p.\ 75, and\hfill\break
in {\bf
Workshop on the Tau/Charm Factory}, Argonne National Laboratory,
June 21--23, 1995, J. Repond, {\it ed.},
AIP Conf.\ Proc.\ No.\ 349, American Institute of
Physics (1996) p.\ 409.

\bibitem{Bigi94}
I. I. Bigi, ``The Expected, the Promised and the Conceivable -- On {\CP}
 Violation
in Beauty and Charm Decays," UND-HEP-94-BIG11, hep-ph/9412227,
to appear in {\sl Proc.\ HQ94
Workshop}, Charlottesville, VA, Oct.\ 7--10, 1994.

\bibitem{left-right}
A. Le Yaouanc {\it et al.}, Phys.\ Lett.\ B {\bf 292}, 353 (1992);\hfill\break
M. Gronau and S. Wakaizumi, Phys.\ Rev.\ Lett.\ {\bf 68}, 1814 (1992).

\bibitem{Pakvasa2000}
S. Pakvasa,
in {\bf The Future of High-Sensitivity Charm
Experiments}, {\it op. cit.}, p.\ 85.

\bibitem{Xing}
Z. Xing, Phys.\ Lett.\ B {\bf 353}, 313 (1995).

\bibitem{CHEOPS}
Yu.\ Alexandrov {\it et al.}, ``CHarm Experiment with Omni-Purpose Setup,"
Letter of Intent to CERN, CERN/SPSLC 95-22, March 28, 1995.

\bibitem{Frabetti}
P. L. Frabetti {\it et al.}, Phys.\ Rev.\ D {\bf 50}, R2953 (1994).

\bibitem{Bartelt}
J. Bartelt {\it et al.} (CLEO Collaboration), 
Phys.\ Rev.\ D {\bf 52}, 4860 (1995).

\bibitem{Aitala}
E. M. Aitala {\it et al.}, Phys.\ Rev.\ Lett.\ {\bf 76}, 364 (1996).

\bibitem{Buccella}
F. Buccella {\it et al.}, Z. Phys.\ C {\bf 55}, 243 (1992).

\bibitem{Purohit-Weiner}
M. Purohit and J. Weiner, ``Preliminary Results on the Decays $D^+\to
K^+\pi^+\pi^-,\,
D^+\to K^+ K^+ K^-$," FERMILAB-Conf-94/408-E, to appear in {\sl Proc.\ DPF
'94}, Albuquerque, NM, Aug.\ 1--8, 1994.

\bibitem{Cinabro}
D. Cinabro {\it et al.} (CLEO collaboration), Phys.\ Rev.\ Lett.\
{\bf 72}, 1406 (1994).

\bibitem{Purohit95}
M. Purohit, ``$D^0-\overline {D^0}$ Mixing Results from E791," to appear in
{\sl Proc.\ LAFEX International School on High Energy Physics (LISHEP95)},
 Rio de Janeiro, Brazil, Feb.\ 
Feb.\ 7--22, 1995.

\bibitem{E691}
J. C. Anjos {\it et al.} (E691 Collaboration), 
Phys.\ Rev.\ Lett.\ {\bf 60}, 1239 (1988).

\bibitem{mixing}
J. F. Donoghue, E. Golowich, B. R. Holstein, and J. Trampetic,
Phys.\ Rev.\ D {\bf 33}, 179 (1986);\hfill\break
 H. Georgi, Phys.\ Lett.\ B {\bf 297}, 353
(1993);\hfill\break 
T. Ohl, G. Ricciardi, and E. H. Simmons, Nucl.\ Phys.\ {\bf B403},
605 (1993).

\bibitem{MkII-mixing}
G. E. Gladding (Mark II collaboration), 
in {\sl Proc. 5th Int.\ Conf.\ on
Physics in Collision}, Autun, France, Jul.\ 3--5, 1985, B. Aubert and L.
Montanet, {\it eds.}, Editions Frontieres (1985), p.\ 259.

\bibitem{Wolfenstein_mixing}
L. Wolfenstein, Phys.\ Lett.\ {\bf 164B}, 170 (1985).

\bibitem{Wolfenstein}
L. Wolfenstein, Phys.\ Rev.\ Lett.\ {\bf 75}, 2460 (1995).

\bibitem{Datta}
A. Datta, Phys.\ Lett.\ {\bf 154B}, 287 (1985).

\bibitem{Hadeed}
A. Hadeed and B. Holdom, Phys.\ Lett.\ {\bf 159B}, 379 (1985).

\bibitem{Bigietal}
I. I. Bigi, F. Gabbiani, and A. Masiero,
Z. Phys.\ C {\bf 48}, 633 (1990).

\bibitem{Bigi-Sanda2}
I. I. Bigi and A. F. Sanda, Phys.\ Lett.\ B {\bf 171}, 320 (1985).

\bibitem{multiple-Higgs}
 L. F. Abbott, P. Sikivie, and M. B. Wise, Phys.\ Rev.\ D {\bf 21}, 1393
(1980);\hfill\break
 V. Barger, J. L. Hewett, and R. J. N. Phillips,
 Phys.\ Rev.\ D {\bf 41}, 3421
(1990);\hfill\break
J. L. Hewett, Phys.\ Rev.\ Lett.\ {\bf 70}, 1045 (1993).

\bibitem{Nir}
Y. Nir and N. Seiberg, Phys.\ Lett.\ B {\bf 309}, 337 (1993).

\bibitem{Techni}
 E. Eichten, I. Hinchliffe, K. D. Lane, and C. Quigg,
Phys.\ Rev.\ D {\bf 34},
1547 (1986).

\bibitem{Lepto}
W. Buchmuller and D. Wyler, Phys.\ Lett.\ {\bf 177B}, 377 (1986) and
Nucl.\ Phys.\ {\bf B268}, 621 (1986);\hfill\break 
Miriam Leurer, Phys.\ Rev.\ Lett.\ {\bf 71}, 1324 (1993).

\bibitem{seesawLR}
A. S. Joshipura, Phys.\ Rev.\ D {\bf 39}, 878 (1989).

\bibitem{Babu}
 K. S. Babu, X. G. He, X. Li, and S. Pakvasa, 
Phys.\ Lett.\ B {\bf 205}, 540 (1988);\hfill\break
T. G. Rizzo, Int.\ J.\ Mod.\ Phys.\ {\bf A4}, 5401 (1989).

\bibitem{Bigi87}
I. I. Bigi, in
{\bf Charm Physics}, {\sl Proc.\ Int.\ Symp.\ on Charm Physics},
Beijing, China, June 4--16, 1987, Gordon and Breach (1987), p.\ 339.

\bibitem{Blaylock}
G. Blaylock, A. Seiden, and Y. Nir, 
Phys.\ Lett.\ B {\bf 355}, 555 (1995).

\bibitem{Browder}
T. E. Browder and S. Pakvasa, ``Experimental Implications of Large {\CP}
Violation and Final State Interactions in the Search for $D^0- \overline {D^0}$
Mixing," UH-511-828-95 (1995).

\bibitem{Liu}
T. Liu,  in {\bf The Future of High-Sensitivity Charm
Experiments}, {\it op cit.}, p.\ 375, and
Ph.D. Thesis, Harvard
University, HUHEPL-20 (1995).

\bibitem{Bigi2000}
I. I. Bigi, in {\bf The
Future of High-Sensitivity Charm Experiments}, {\it op cit.}, p.\ 323.

\end{thebibliography}
\end{document}